\documentclass[preprint, 3p]{elsarticle}

\usepackage{graphicx}
\usepackage{amssymb}
\usepackage{amsmath}
\usepackage[section]{placeins}

\setcitestyle{authoryear}
\setcitestyle{round}
\setcitestyle{semicolon}

\usepackage{lineno}
\usepackage{url}
\usepackage[]{algorithm2e}
\usepackage{setspace}

\usepackage[dvipsnames]{xcolor}
\makeatletter
\def\ps@pprintTitle{%
 \let\@oddhead\@empty
 \let\@evenhead\@empty
 \def\@oddfoot{}%
 \let\@evenfoot\@oddfoot}
\makeatother

\journal{Computational Statistics \& Data Analysis}

\begin{document}

\let\today\relax

\begin{frontmatter}

\title{SCADDA: Spatio-temporal cluster analysis with density-based distance augmentation and its application to fire carbon emissions}

\author[first,second]{Ben Moews\corref{corresponding}}
\ead{ben.moews@ed.ac.uk}
\cortext[corresponding]{Corresponding author}

\author[first]{Antonia Gieschen}
\ead{antonia.gieschen@ed.ac.uk}

\address[first]{Business School, University of Edinburgh, 29 Buccleugh Pl, Edinburgh, EH8 9JS, UK}
\address[second]{Centre for Statistics, University of Edinburgh, Peter Guthrie Tait Rd, Edinburgh, EH9 3FD, UK}

\date{}

\begin{abstract} 
Spatio-temporal clustering occupies an established role in various fields dealing with geospatial analysis, spanning from healthcare analysis to environmental science. One major challenge are applications in which cluster assignments are dependent on local densities, meaning that higher-density areas should be treated more strictly for spatial clustering and vice versa. Meeting this need, we describe and implement an extended method that covers continuous and adaptive distance rescaling based on kernel density estimates and the orthodromic metric, as well as the distance between time series via dynamic time warping. In doing so, we provide the wider research community, as well as practitioners, with a novel approach to solve an existing challenge as well as an easy-to-handle and robust open-source software tool. The resulting implementation is highly customizable to suit different application cases, and we verify and test the latter on both an idealized scenario and the recreation of prior work on broadband antibiotics prescriptions in Scotland to demonstrate well-behaved comparative performance. Following this, we apply our approach to fire emissions in Sub-Saharan Africa using data from Earth-observing satellites, and show our implementation's ability to uncover seasonality shifts in carbon emissions of subgroups as a result of time series-driven cluster splits.
\end{abstract}


\end{frontmatter}

\nolinenumbers 



\section{Introduction}
\label{sec:introduction}

Understanding similarities in patterns is a useful approach to the analysis of various data types stemming from a range of sources. These can include, to name a few, environmental, social, and natural sciences. Grouping observations in terms of similarity or distance allows researchers to make statements and decisions based on the behavior of objects within groups as well as differences between them. This provides benefits over treating all observations the same or developing a large number of different approaches separately, for example by detecting trends and shifts in behaviours which can be addressed through targeted interventions.

One approach to this segmentation approach is clustering, which allows for the grouping of data points into groups (clusters) without manually pre-labelling observations using known classifications \citep{Wunsch2008}. Due to a growing interest in collecting complex data, driven by better storage and computation abilities, spatio-temporal data has also been a point of increased interest with regard to specialized analysis approaches \citep{kisilevich2010}. Considering the wide-spanning importance of these approaches, they should be accessible to a broad range of researchers from different backgrounds.

Spatio-temporal analysis in particular can be defined as the grouping of elements according to both their spatial closeness and behavior over time, which can be a complex endeavor especially for practitioners. Available methods are often complicated in their implementation and not tailored specifically to real-life data scenarios, as opposed to simulations and theoretical settings.

Regardless of these challenges, there has been growing interest in the use of spatio-temporal data, particularly in the area of environmental science, for example in the prediction of fires or flooding, highlighting the usefulness of different approaches from the data mining and machine learning literature \citep{cheng2008, Tonini2022}.

In this context, cluster analysis is particularly useful for applications that lack prior knowledge about the patterns in the data. There exist a large number of algorithms for this purpose, with the choice of method usually being driven by the given context as well as the available data. One way of categorising clustering approaches is by sorting them into partitioning, hierarchical, graph-based, probabilistic and density-based methods, depending on the cluster-forming process.

Partitioning algorithms include, for example, $k$-means, which separates the data space into a number $k$ of clusters by randomly selecting cluster centers (centroids) to which the nearest observations are then assigned. The result is then iteratively improved by relocating these centroids to optimize the within-cluster sum of squares \citep[see, for example,][for a review of $k$-means and its recent applications]{ahmed2020}. Within environmental applications, \citet{gorsevski2003} present an application of a $k$-means-based method to analyse the occurrence of landslides, while \citet{xu2018} use it to create a map of flood risks.

Hierarchical clustering algorithms include both agglomerative and divisive approaches. The former partition the dataset by starting with each observation assigned to its own cluster and then step-wise merging the most similar clusters until all observations are assigned the same label, while the latter do the opposite \citep[see, for example,][for a recent review]{tokuda2022}. Keeping within the same area of application, \citet{pappada2018} apply agglomerative clustering to analyze concurrent flood risks, which includes copulas to model spatial dependency between the locations.

Graph-based algorithms utilize concepts from graph theory to create cluster sub-trees by optimising the best cut to divide the dataset, as seen for example in \citet{yang2019}, who utilise graph-based clustering to detect spatio-temporal patterns in the use of rental bikes.

Probabilistic clustering, on the other hand, makes use of techniques such as mixture models to identify sub-populations in the data space which are then denoted as clusters. \citet{sun2019} demonstrate that by using Gaussian mixture models in combination with copulas to model spatio-temporal dependency, they are able to improve forecasting of wind power in locations of wind farms and reduce computational costs due to the use of clustering.

For further details on these methods, the interested reader is referred to \citet{Wunsch2008} for an overview of clustering approaches, as the remainder of this paper focuses on density-based methods. This group of algorithms uses concepts of point density for their definition of clusters, which makes them intuitive for practitioners and especially suitable for spatial data \citep{Moayedi2019}.

(Geo-)spatial data describes information typically observed within a two-dimensional geographic area. Within this broad field of application, one focal point of interest in many cases is the grouping of data points according to their geospatial closeness, sometimes in combination with their similarity in some other dimension. Spatial clustering as the detection of patterns within spatial data is an approach which is used in many different areas of research, such as environmental and social sciences, as it allows for this segmentation of spatial data points without the need for pre-labeling them \citep{Baghbanan2020, Park2020, Lurka2021}. Due to the spatial nature, it also allows for the identification and addressing of localized problems specific to an area of interest, for example by policy and decision-makers.

Density-Based Spatial Clustering of Applications with Noise (DBSCAN) in particular has found success in spatial clustering applications \citep{Ester1996}. DBSCAN utilizes concepts of point density to detect and label groups of points which are close to each other while being separated from other such groups. This algorithms has been used in many different areas of social science research, such as urban traffic patterns and tourism behavior, as well as the social media behavior of consumers related to food trends \citep{Toshniwal2020, Munoz2019, Hopken2020, Pindado2020, Toshniwal2020}.

However, a known issue of the algorithm is the detection of clusters in spaces with varying point densities. The reason for this weakness lies in the definition of density, which does not differentiate between generally non-dense (sparse) and dense areas. As a result, what would qualify as a cluster in a sparse area might not reach the user-defined threshold of density (and thus cluster existence) in a nearby dense space.

This has been acknowledged by some of the the method's original authors as well as other related works in the literature, which are offering their respective solutions \citep{Kriegel2011}. These generally employ neighbor-based approaches such as $k$-nearest neighbor to identify the local point density \citep[see, for example,][]{Biccici2007, Ertoz2003,Pei2009} or the use of a local scaling or density factor \citep{Zelnik2004, Birant2007}. 

The other half of spatio-temporal clustering, as the name suggests, deals with time series, meaning sequential strings of observations which, in the case of real-world data, were collected over time at specified time intervals. Temporal clustering assigns cluster labels based on the similarity of these series \citep{caiado2015}. This can, for example, be achieved by comparing the elements of two series directly (lock-step or observation-based approaches), calculating a similarity measure based on a number of time series features (feature or model-based approaches), or considering the overall shape of the time series (shape-based or elastic approaches) \citep{caiado2015}.

The choice of approach depends on different factors, for example whether the time series are of equal length as well as the cost of calculating dissimilarity matrices, which depends on the series length and number of observations. Another potential issue which affecting the decision is the challenge of shifted or warped time series, where the overall shape of two observations' sequences are similar but exhibit those similarities at different points in time. An overview and comparison of different approaches can be found in \citet{Aghabozorgi2015}, while an empirical comparison of different dissimilarity measures used for time series clustering is provided by \citet{Serra2014}.

If then, in addition to temporal data, the given dataset is also described through spatial locations at which the temporal data is occurring or being collected, this dataset is called spatio-temporal. Cluster analysis for spatio-temporal data can be very complex, but has proven useful in different applications such as the occurrence of natural disasters in different locations over time as well as the detection and spread of diseases \citep{Sugumaran2009, Wu2017, Tonini2022}. An overview of spatio-temporal clustering can be found in \citet{kisilevich2010}, who define methods depending on the data format of both the temporal and spatial data dimensions. Spatial locations can be considered as either fixed or dynamic points, while temporal data can be considered as single snapshots, updated snapshots, or time series of the same object. 

Methods for spatio-temporal clustering usually combine approaches from both spatial and temporal approaches by simultaneously considering closeness in terms of spatial location and time series similarity. This can, for example, be seen in the ST-DBSCAN algorithm, which forms part of this work and is described in more detail in Section~\ref{subsec:stdbscan} \citep{Birant2007}.

Two challenges in spatial and temporal clustering are addressed through this paper: First, varying point density has been identified as a challenge for spatial clustering algorithms including DBSCAN, as it leads to the method not being able to detect clusters in both dense and sparse regions. What qualifies as a cluster in a sparse region does not necessarily qualify as one in dense areas of the data space, although the detection of both might be important for a given application case. \citet{Gieschen2022} introduce an adaptation of the ST-DBSCAN algorithm that uses a kernel density-based approach to modifying spatial dissimilarity matrices to make dense and sparse areas more comparable.

Secondly, a known problem in time series clustering is the comparison of shifted or warped time series, which exhibit the same underlying shape (and thus, behaviour) but cannot be detected with lock-step or similar dissimilarity-based methods. The same authors make use of dynamic time warping (DTW) to take this into consideration and allow for the identification of time series of similar overall shape.

The main motivation of this paper is, therefore, to employ the previous empirical work by \citet{Gieschen2022}, which presents an introduction to a spatio-temporal clustering approach addressing both of these challenges. We extend this approach through additional functionalities such as outlier-resistant $z$-score normalization, maximum percentages of non-assigned data points through secondary clusters, and reduced computational costs through constrained dynamic time warping. In doing so, we provide researchers with the first open-source implementation of this algorithm for spatio-temporal cluster detection in the form of an accessible and easy-to-use software package. The focus lies on the applicability of the method in cases of uneven data density distributions, while taking shifted and warped time series into consideration.

By providing an open-source software tool, researchers from different disciplines and backgrounds are able to apply our general-purpose module, improving transparency and replicability. We recreate the empirical findings in \citet{Gieschen2022} with the same data to test our implementation, and validate the spatial rescaling in a simulation comparison. Based on our application to fire carbon emissions, we also provide a tool for emerging challenges in the light of climate change increasing the risk of devastating wildfires.

The remainder of this paper is structured as follows. In Section~\ref{sec:methodology}, we cover the ST-DBSCAN algorithm as the base for our work, provide an in-depth description of our density-based distance rescaling approach, and introduce dynamic time warping and curvature-aware distance metrics. Following this, Section~\ref{sec:implementation} introduces the required and optional inputs and parameters, as well as our algorithm, its pseudocode, and installation instructions for researchers and practitioners. Our experiments are summarized in Section~\ref{sec:experiments}, going from an idealized test scenario to the replication of healthcare analysis results and an application to carbon emission from fires. Lastly, we discuss our method's general applicability, advantages as well as limitations, and suggestions for further research in Section~\ref{sec:discussion}, and provide our conclusions in Section~\ref{sec:conclusion}.


\section{Methodology}
\label{sec:methodology}

\subsection{ST-DBSCAN for spatio-temporal analysis}
\label{subsec:stdbscan}

\begin{figure}[!htb]
    \centering
    \includegraphics[width=\textwidth]{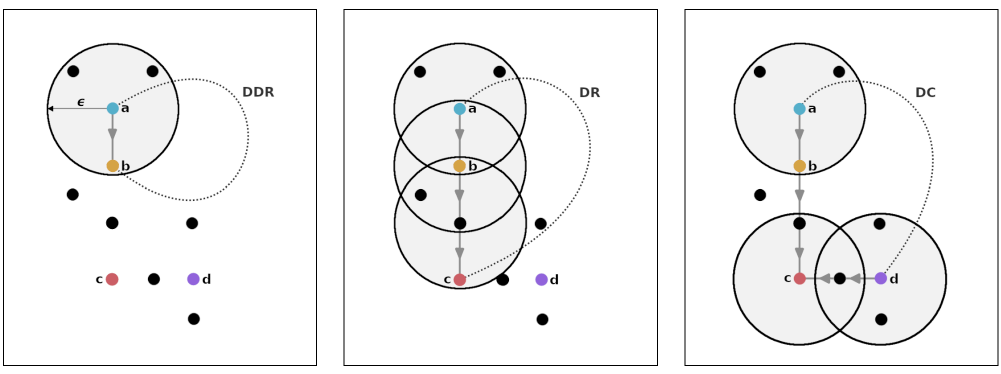}
    \caption{Visualization of reachability concepts in DBSCAN with $\lambda = 3$. The left panel shows point $b$ being directly density-reachable (DDR) from core point $a$ with $| N_\epsilon(a) | \geq 3$ and $b \in N_\epsilon(a)$, while the center panel depicts point $c$ being density-reachable (DR) through a chain of two DDR-linked core points. Lastly, the right panel shows that point $d$ is density-connected (DC) to point $a$ through point $c$, with the latter being DR-linked to points $a$ and $d$.}
    \label{fig:dbscan}
\end{figure}

As discussed in Section~\ref{sec:introduction}, DBSCAN is a clustering algorithm using concepts of distance and neighborhood to identify areas of high point density, which are then classified as clusters. In this context, the $\epsilon$-neighborhood $N_{\epsilon}(x)$ of a data point $x$ is defined as
\begin{equation}
N_{\epsilon}(x) = \{ y \in B \ | \ D(x,y)\leq \epsilon \},
\label{eq:dbscan_neighborhood}
\end{equation}
where $x$ and $y$ are two points in a database $B$ whose distance is denoted as $D(x,y)$, and $\epsilon$ is a chosen parameter denoting the maximum radius accepted for the formation of a cluster. In order to define the minimum number of points required in a neighborhood to classify the area as a cluster, \citet{Ester1996} introduce the concepts of reachability and connectedness in this context. Here, a point $y$ is \textit{directly density-reachable} from $x$ if
\begin{equation}
y \in N_{\epsilon}(x), \ \mathrm{with} \ |N_{\epsilon}(x)| \geq \lambda,
\label{eq:DDR-dbscan}
\end{equation}
where $\lambda$ is a chosen parameter defining the minimum number of points needed in a neighborhood to form a cluster. This definition is fulfilled for pairs of core points, meaning data points located in the middle of a cluster, but not necessarily for pairs of border points, meaning points located toward the edge of a cluster, as illustrated in the left panel of Figure~\ref{fig:dbscan}. This concept thus creates clusters in areas of high point density.

A point $y$ is further \textit{density-reachable} from point $x$ if there exists a `chain' of $t$ data points, $\{ p_1, p_2, \dots , p_t \}$, with $p_1 = x$ and $p_t = y$, which are directly density-reachable from each other. This condition allows for the formation of arbitrarily shaped clusters by creating a linking effect shown in the center panel of Figure~\ref{fig:dbscan}. Lastly, points $x$ and $y$ are \textit{density-connected} if there exists an additional point $o$ which is density-reachable from both $x$ and $y$, as demonstrated in the right panel of Figure~\ref{fig:dbscan}.

\citet{Ester1996} now define a cluster $C$ as a non-empty subset of a database $B$ which fulfills the two following conditions: First, for $x, y \in B$, if $x \in C$ and $y$ is density-reachable from $x$, then $y \in C$. Secondly, for any pair $x, y \in C$, $x$ is density-connected to $y$ with regard to $\epsilon$ and $\lambda$. Lastly, for a set of $k$ clusters, $\{ C_1, C_2, \dots ,C_k \},$ with regard to parameters $\epsilon_i$ and $\lambda_i$ with $i \in \{1, 2, \dots ,k \}$, any point $x \in B \ | \ \forall i: x \notin C_i$ is denoted as \textit{noise}, that is, an outlier which is not part of any cluster.


The spatio-temporal adaption of DBSCAN, aptly coined ST-DBSCAN by \citet{Birant2007}, extends the above concept to additional time series data. For a point $x$, instead of one neighborhood $N_{\epsilon}(x)$, we extend Equation \ref{eq:dbscan_neighborhood} to define two neighborhoods, $N_{\epsilon_s}(x)$ and $N_{\epsilon_t}(x)$, which refer to the spatial and temporal neighborhood, respectively, and which can be written as
\begin{equation}
\begin{split}
N_{\epsilon_s}(x) &= \{y \in B \ | \ D(x,y) \leq \epsilon_s\},\\
N_{\epsilon_t}(x) &= \{y \in B \ | \ D(x,y) \leq \epsilon_t\}.
\end{split} 
\end{equation} 
Subsequently, we extend Equation \ref{eq:DDR-dbscan} and define $y$ to be directly density-reachable from $x$ if
\begin{equation}
    \begin{split}
         y \in N_{\epsilon_s}(x),\\
         y \in N_{\epsilon_t}(x),\\
         |N_{\epsilon_s}(x)| \geq \lambda,\\
         |N_{\epsilon_t}(x)| \geq \lambda.
    \end{split}
\label{eq:DDR-stdbscan}
\end{equation}

Following these definitions, it becomes apparent that ST-DBSCAN allows for the formation of clusters only in cases in which points $x$ and $y$ are in both spatial and temporal neighborhoods of each other simultaneously, effectively adding a further temporal constraint on cluster identification. Furthermore, these neighborhoods both have to comply with the requirement of having at least $\lambda$ neighbors to form a cluster.


\subsection{Density-based distance matrix rescaling}
\label{subsec:distance_rescaling}

Following the motivation in Section~\ref{sec:introduction}, we are now faced with the need to implement a distance-based rescaling approach. This will allow us to use global parameters, for example for the maximum distance of neighbors to be considered part of the same cluster, when investigating areas of application that deal with large density discrepancies due to factors extraneous to a given topic of interest.

To get a density-based landscape overlaying the spatial data distribution, we make use of kernel density estimation (KDE). Pioneered by \citet{Rosenblatt1956} and \citet{Parzen1962}, this approach places, in simple terms, a chosen distribution over each data point as a smoothing kernel. More formally, it can be written as
\begin{equation}
\mathrm{KDE} (x, \beta) = \frac{1}{|\theta|} \sum\limits_{i = 1}^{|\theta|} \frac{1}{\beta^d} \mathcal{K} \left( \frac{|| x - \theta_i ||}{\beta} \right),
\label{eq:kde_basis}
\end{equation}
for a given data point $x$, a dataset $\theta$, a bandwidth $\beta$, and a smoothing kernel $\mathcal{K}$, and with $d$ denoting the dimensionality of the problem \citep{Bishop2006}. Such kernels generally have to fulfill two requirements,
\begin{equation}
    \begin{split}
       \mathcal{K}(\Phi) &\leq 0,\\
       \int \mathcal{K} (\Phi) \mathrm{d} \Phi &= 1 ,
    \end{split}
\end{equation}
for given parameters $\Phi$, guaranteeing that probability distributions that follow from the kernel are non-negative everywhere and integrate to unity. The Gaussian kernel is a common choice and has the form
\begin{equation}
\mathcal{K}_\mathrm{Gaussian} (x, \beta) = \frac{1}{\sqrt{2 \pi} \beta} e^{- \frac{x^2}{2 \beta^2}}.
\end{equation}
In choosing the Gaussian kernel for Equation~\ref{eq:kde_basis}, the latter can then be rephrased as
\begin{equation}
\mathrm{KDE}_{\mathrm{Gaussian}} (x, \beta) = \frac{1}{|\theta|} \sum\limits_{i = 1}^{|\theta|} \frac{1}{(2 \pi \beta^2)^{\frac{1}{2}}} e^{- \frac{|| x - \theta_i ||^2}{2 \beta^2}},
\label{eq:kde}
\end{equation}
where $\beta$ takes the place of the kernel's standard deviation. For the data-driven bandwidth determination, \citet{Gieschen2022} apply Scott's rule, as introduced by \citet{Scott1992},
\begin{equation}
\hat{\beta} = |\theta|^{- \frac{1}{d + 4}}.
\label{eq:scott}
\end{equation}
An alternative is provided by \citet{Silverman1986}, with both rules of thumb being wide-spread in the statistical and domain application literature, and takes the form
\begin{equation}
\hat{\beta} = \left( \frac{| \theta | \cdot (d + 2)}{4} \right)^{- \frac{1}{d + 4}},
\label{eq:silverman}
\end{equation}
These are also the most common methods in a variety of major statistical libraries, and are attractive primarily due to their broad suitability for reasonable estimates when encountering real-world datasets, as well as their extremely low computational cost. However, should a more involved bandwidth optimization method be desired despite considerably increases in runtime, this option should be available, which is why we opt for a choice between both rules and a directly provided numerical value in Section~\ref{subsec:input_parameters}.

For the logistic distance rescaling, let $\overline{g}$ be the density average as measured by Equation~\ref{eq:kde} for a given dataset $\theta$, and $g_i$ the separate densities for data points with $i \in (1, 2, \dots , |\theta|)$. In order to rescale spatial distances depending on the local data point density, we implement a logistic function of the general form
\begin{equation}
f(x) = \frac{f_\mathrm{max}}{1 + e^{- k (x - x_0)}},
\label{eq:logistic_basis}
\end{equation}
where $f_\mathrm{max}$ denotes the sigmoid function's maximum value, $x_0$ the function's half-way value on the horizontal axis, and $k$ the curve's steepness. In keeping with the base methodology of \citet{Gieschen2022}, we then transform kernel density estimates into rescaling weights.

This means that we compress and stretch distances in a continuous fashion for low-density and high-density areas, respectively, akin to morphing the coordinate space in a data-driven way. Accordingly, let $x_0 = \overline{g}$ and $f_\mathrm{max} = 2$ to center the function around an effectless multiplication weight of one, 
\begin{equation}
f(g_i) = \frac{2}{1 + e^{- k (g_i - \overline{g})}},
\label{eq:logistic}
\end{equation}
thus setting an upper limit of double the pre-rescaling distance. While the lower limit is zero, which would shift a distance to that value, this only applies in case of two data points with exactly the same spatial coordinates, in which case the pre-rescaling distance is already zero and nothing changes.

This choice of function has the advantage of an adjustable steepness, which translates to the potential to alter the strength of the density-based rescaling, offering an easy way to customize the behavior to specific challenges. It also furthers the continuous approach of using kernel density estimates to approximate the distribution of spatial data points. For any two spatial data points, the rescaled distance $\hat{s}_{i, j}$ is then
\begin{equation}
\hat{s}_{i, j} = s_{i, j} \cdot \left( \frac{f(g_i) + f(g_j)}{2} \right) = s_{i, j} \cdot \left( \frac{2}{1 + e^{- k (g_i - \overline{g})}} + \frac{2}{1 + e^{- k (g_j - \overline{g})}} \right) \cdot 2^{-1},
\end{equation}
using Equation~\ref{eq:logistic_basis}, with the term in brackets serving as the multiplicative rescaling weight.

\subsection{Dynamic time warping and distances}
\label{subsec:dtw}

Dynamic time warping (DTW) is an approach to similarity measurement in time series analysis, comparing two separate temporal sequences. Most notably, it allows for a lag, meaning that isolated effects influencing the position of structural features in one time series but not another are accounted for. This is especially useful when dealing with real-world data with extraneous influences that are locally restricted, as confounding variables can present a major challenge in some research areas, for example in the context of lagged onsets of health-related correlations due to unequal age distributions.

For an explanation of DTW, let $x = (x_1, x_2, \dots , x_n)$ be a real-values time series with $x_i \in \mathbb{R}$, and $\nu = (\nu_1, \nu_2, \dots , \nu_p)$ a corresponding path of order $m \times n$.  We can then write the boundary conditions as 
\begin{equation}
    \begin{split}
        \nu_1 &= (1, 1), \\
        \nu_p &= (m, n),
    \end{split}
\end{equation}
while the step condition follows the form
\begin{equation}
\forall \rho \in [p - 1] : \nu_{\rho + 1} - \nu_\rho \in \{ (1, 0), (0, 1), (1, 1) \},
\end{equation}
for a number $p$ of points in $\nu_\rho = (i_\rho, j_\rho) \in [m] \times [n]$. Let $\mathcal{A}_{m, n}$ be the set of admissible paths of the described order, then $\nu \in \mathcal{A}_{m, n}$ represents a path $\nu$ as an expansion of two given series $x = (x_1, x_2, \dots , x_m)$ and $y = (y_1, y_2, \dots , y_n)$, to $\wp_\nu (x) = (x_{i_1}, x_{i_2}, \dots x_{i_p})$ and $\eth_\nu (y) = (y_{i_1}, y_{i_2}, \dots y_{i_p})$ \citep[see][for a more detailed introduction]{Jain2019}. With a cost of transforming $x$ and $y$ toward $\nu$ given by
\begin{equation}
    \mathcal{T}_\nu (x, y) = || \wp_\nu (x) - \eth_\nu (y) ||^2 = \sum_{(i, j) \in \nu} (x_i - y_i)^2,
\end{equation}
the DTW measure can then be written as
\begin{equation}
    \mathrm{DTW} (x, y) = \min \left( \sqrt{\mathcal{T}_\nu (x, y)} \right).
\label{eq:dtw}
\end{equation}
It should be noted that, while the latter is distance-like, it does not technically constitute a distance metric in terms of the triangle equality, meaning
\begin{equation}
    \mathrm{DTW}(\mathbf{x} + \mathbf{y}) \leq \mathrm{DTW}(\mathbf{x}) + \mathrm{DTW}(\mathbf{y}),
\end{equation}
as the latter is not guaranteed to hold. As an easily accessible example, this is the case for time series $\upsilon_1 = (0, 1, 1, 2)$, $\upsilon_2 = (0, 1, 2)$, and $\upsilon_3 = (0, 2, 2)$, for which the resulting measurements are $\mathrm{DTW}(\upsilon_1, \upsilon_3) = 2$,  $\mathrm{DTW}(\upsilon_1, \upsilon_2) = 0$, and  $\mathrm{DTW}(\upsilon_2, \upsilon_3) = 1$. Computational expense is, of course, another concern when dealing with applications to datasets. The Sakoe-Chiba band instroduced by \citet{Sakoe1978} places a constraint on the path described above, with the goal to eliminate reduncant computations.

The concept is simple; the distance matrix in which the path is situated is constrained by enforcing $|| d_i - d_j || \leq \xi$ for a window size $\xi$ and two matrix positions $d_i$ and $d_j$. This approach creates a `band' around the matrix diagonal, reducing the computational complexity to $\mathcal{O}(\xi (n + m))$.

With temporal distances measurements taken care of, the next question concerns the spatial distances between data points. In the case of geospatial analysis, this commonly means that each data point in a dataset $\theta$ presents a tuple $(\varphi_i, \varrho_i)$ for $i \in (1, 2, \dots , |\theta|)$, with $\varphi$ and $\varrho$ representing latitude and longitude, respectively. When operating on small scales relative to the curvature of the Earth, the Euclidean distance,
\begin{equation}
    D_\mathrm{Euclidean} = \sqrt{(\varphi_2 - \varphi_1)^2 + (\varrho_2 - \varrho_1)^2},
\label{eq:euclidean}
\end{equation}
offers a sufficient approximation by treating the globe's surface as locally flat for small-enough areas. This is reasonable as long as a method's potential application cases are limited to such areas, as local geography is likely to have a larger impact than the planet's shape. However, as soon as such a method is applied to larger distances, for example when performing geospatial analyses on data points spanning multiple countries, this presents a problem. For this reason, the prior work by \citet{Gieschen2022}, which lead to the development of SCADDA, uses the following way to calculate the distance between latitude-longitude coordinates,
\begin{equation}
    \begin{split}
        D_{\mathrm{Meeus}} &= 2\cdot \arctan{\sqrt{\frac{S}{C}}} \cdot \vartheta (f H_1 \sin^2 \frac{\varphi_1 + \varphi_2}{2} \cos^2 \frac{\varphi_1 - \varphi_2}{2} - f H_2 \cos^2 \frac{\varphi_1 + \varphi_2}{2} \sin^2 \frac{\varphi_1 - \varphi_2}{2}), \ \mathrm{with}\\
        S &= \sin^2 \frac{\varphi_1 - \varphi_2}{2} \cos^2 \frac{\varrho_1 - \varrho_2}{2} + \cos^2 \frac{\varphi_1 + \varphi_2}{2} \sin^2 \frac{\varrho_1 - \varrho_2}{2}, \ \ H_1 = \frac{3 R - 1}{2 C},\\
        C &= \cos^2 \frac{\varphi_1 - \varphi_2}{2} \cos^2 \frac{\varrho_1 - \varrho_2}{2} + \sin^2 \frac{\varphi_1 + \varphi_2}{2} \sin^2 \frac{\varrho_1 - \varrho_2}{2}, \ \ H_2 = \frac{3 R + 1}{2 S}, \ \mathrm{and}\\
        R &= \frac{\sqrt{S C}}{\arctan(\sqrt{\frac{S}{C}})}.
    \end{split}
\label{eq:meeus}
\end{equation}
Here, $\vartheta$ denotes the equatorial radius, and the relative error has the order of the square of the planet's flattening $f$. This formula is often named after its popularization in astronomical research by \citet{Meeus1991}, and accurate to round-off and guaranteed to converge, but traces back to work publicized in the early 1950s by the Bureau des Longitudes, a long-established French scientific institution for the improvement of navigation, time standardization, geodesy, and observational astronomy. While reasonably accurate, the main disadvantage is the computational cost due to taking the obloid shape of the Earth into account\footnote{An ellipsoid of revolution is, of course, also a mathematical approximation to the exact shape of the planet. In this context, it is worth mentioning that other shapes have been proposed, notably a pear-shaped Earth by Christopher Columbus due to his blatant inability to correctly measure the diurnal motion of the North Star. We are not aware of a distance metric that incorporates this assumption, but it would certainly be a suitable topic for the tradition of April 1 arXiv papers.}, which is why orthodromic distance metrics, also known as the great-circle distance, are commonly chosen.

Conversely, the haversine function has a similarly long standing in the study and practice of navigation \citep[see][]{Inman1835}, and is numerically better-conditioned for small distances and robust to large rounding errors for the latter. Frequently used on small scales in geospatial analysis, for example by \citet{Moews2021} for city-level calculations while maintaining large-distance accuracy, it can be written as
\begin{equation}
    \begin{split}
        D_\mathrm{Haversine} &= 2 r \cdot \arcsin \left( \sqrt{\mathrm{hav} (\varphi_2 - \varphi_1) + (1 - \mathrm{hav} (\varphi_1 - \varphi_2) - \mathrm{hav} (\varphi_1 + \varphi_2) \cdot \mathrm{hav}(\varrho_2 - \varrho_1))} \right)\\
         &= 2 r \cdot \arcsin \left( \sqrt{\sin^2 \left( \frac{\varphi_2 - \varphi_1}{2} \right) + \cos \varphi_1 \cdot \cos \varphi_2 \cdot \sin^2 \left( \frac{\varrho_2 - \varrho_1}{2} \right)} \right), \ \mathrm{with}\\
         \mathrm{hav}(x) &= \sin^2 \frac{x}{2} = \frac{1 - \cos(x)}{2}.
    \end{split}
\end{equation}
For the purpose of our implementation, however, we want to avoid the potential for rounding errors when encountering antipodal points. We thus opt for the more modern special case of the Vincenty formula,
\begin{equation}
    D_\mathrm{Vincenty} = \arctan \frac{\sqrt{(\cos \varphi_2 \sin (|| \varrho_2 - \varrho_1 ||))^2 + (\cos \varphi_1 \sin \varphi_2 - \sin \varphi_1 \cos \varphi_2 \cos (|| \varrho_2 - \varrho_1 ||))^2}}{\sin \varphi_1 \sin \varphi_2 + \cos \varphi_1 \cos \varphi_2 \cos (|| \varrho_2 - \varrho_1 ||)},
\label{eq:vincenty}
\end{equation}
which is accurate for all distances and maintains the simplicity of the orthodromic distance by assuming a spheroid with equal major and minor axes. Initially proposed by \citet{Vincenty1975} for the general case, it was later reformulated by \citet{Karney2013} and is now a staple in geostatistical software. While offering a much faster runtime than Equation~\ref{eq:meeus} that is more suitable for practical applications, the Euclidean distance in Equation~\ref{eq:euclidean} is sufficient for small distances over which local geography has a larger effect, thus motivating the option to choose either metric depending on the requirements of a given research application.


\section{Implementation}
\label{sec:implementation}

\subsection{Input parameters and computations}
\label{subsec:input_parameters}

The algorithm requires six inputs to be set in advance. The first is a spatial dataset ($\mathbf{s}$) as an $N \times 2$ matrix, for $N$ data points and with the first and second column containing latitude and longitude values, respectively. The second is the temporal dataset ($\mathbf{t}$), which contains a time series per spatial data point, as an $N \times M$ array, with $N$ as above and $M$ as the number of steps in the time series.

The third and fourth parameter are the maximum distances for the spatial ($\epsilon_s$) and temporal ($\epsilon_t$) datasets to consider a given spatial coordinate or time series still part of the same cluster, which needs to be fulfilled for both the spatial and temporal dimension as described in Section~\ref{subsec:stdbscan}.

Lastly, the minimum number of neighbors to constitute a separate cluster, as well as the steepness of the logistic function ($k$) for distance rescaling covered in Section~\ref{subsec:distance_rescaling}, are necessary to tune the algorithm to a given application case. These inputs and parameters, as well as optional ones, are also listed in Table~\ref{tab:variables}, together with the variable names used in the software implementation.

\begin{table}[!h]
\label{tab:variables}
\caption{Input variables of SCADDA. The first column lists variable designations as used in Algorithms~\ref{alg:cluster}--\ref{alg:nb}, the second column shows the corresponding parameter names of the implementation, and the third column provides a description for each entry.}
\begin{center}
\begin{tabular}{llll}
\hline\noalign{\smallskip}
Variable & Parameter & Description & Type \\
\noalign{\smallskip}\hline\noalign{\smallskip}
$\mathbf{s}$ & \texttt{s\_data} & Spatial latitude and longitude coordinates & \texttt{array-like} \\
$\mathbf{t}$ & \texttt{t\_data} & Time series inputs as vectors per data point & \texttt{array-like} \\
$\epsilon_s$ & \texttt{s\_limit} & Maximum for intra-cluster spatial distances & \texttt{integer, float} \\
$\epsilon_t$ & \texttt{t\_limit} & Maximum for intra-cluster temporal distances & \texttt{integer, float} \\
$\lambda$ & \texttt{minimum\_neighbors} & Minimum of neighbors for non-outlier status & \texttt{integer} \\
$k$ & \texttt{steepness} & Curve steepness for logistic distance rescaling & \texttt{integer, float} \\
$\xi$ & \texttt{window\_param} & Window size used for the Sakoe-Chiba band & \texttt{integer, float} \\
$D$ & \texttt{distance\_measure} & Distance metric used for spatial calculations & \texttt{string} \\
$o$ & \texttt{outlier\_perc} & Maximum percentage for outlier assignments & \texttt{integer, float} \\
$z$ & \texttt{z\_score} & Indicator for the use of z-score normalization & \texttt{boolean} \\
$A$ & \texttt{algorithm} & Indicator for SCADDA versus ST-DBSCAN & \texttt{string} \\
$\hat{\beta}$ & \texttt{bandwidth} & Bandwidth or bandwidth estimation method & \texttt{string, float} \\
\noalign{\smallskip}\hline
\end{tabular}
\end{center}
\end{table}

Six additional optional parameters are automatically set to suitable defaults, but can be set manually as well. The first is the window size for the Sakoe-Chiba band introduced in Section~\ref{subsec:dtw} ($\xi$), which is set to $N \cdot 10^{-1}$ by default. The second is the spatial distance metric of choice, which is set to the orthodromic distance in Equation~\ref{eq:vincenty} if no input is provided, and can otherwise be changed to the Euclidean distance in Equation~\ref{eq:euclidean} for a simple and faster computation.

The third is the maximum outlier percentage allowable in the final cluster assignments ($o$). By default, this is set to 100\% to entail the entire dataset and allow for an unimpeded iteration run of the algorithm. The provision of a value for the parameter means that the algorithm will re-run iterations over remaining outliers to assign them to pseudo-clusters until the percentual threshold is met. In practice, this means that the for-loop in Algorithm~\ref{alg:cluster} is iteratively repeated with remaining outliers. In order to ensure convergence, $\epsilon_s$ and  $\epsilon_t$ are doubled with each iteration. This option should be handled with care, as a requirement of few outliers for some applications is counterbalanced by the potential conflation of pseudo-clusters with statistically well-motivated cluster assignments resulting from the initial single-pass run.

The fourth parameter indicates whether to apply z-normalization to the time series $\mathbf{t}$, forcing a zero mean and standard deviation of one ($z$), which is not done by default. The fifth is an indicator on whether to to use SCADDA or the base ST-DBSCAN algorithm for the spatial clustering component ($A$), with the former as the default choice. Finally, the sixth and last optional parameter indicates the use of Scott's versus Silverman's rule in Equations~\ref{eq:scott} and~\ref{eq:silverman}, or alternatively a direct bandwidth value ($\hat{\beta}$). 

\subsection{Algorithm design and implementation}
\label{subsec:algorithm_design}

Now that we have covered the methodological background, our extensions and modifications, and the data types and parameters, we can begin top put everything together. Algorithms~\ref{alg:cluster}--\ref{alg:nb} show the pseudocode for our method in a compact format, using the variables in Table~\ref{tab:variables}. In order to ensure a suitable brevity of the pseudocode, we drop optional parameters by fixing the distance metric $D$ as the orthodromic distance in Equation~\ref{eq:vincenty} and SCADDA as the desired algorithm, assuming no z-score normalization, and omitting the iteration for outlier percentages described in Section~\ref{subsec:input_parameters}.

The pseudocode also drops non-essential functionalities such as the automatic check whether all inputs are of the correct data type and form, as well as consistencies like the number of time series being the same as the number of provided data points, which ensure that no runtime errors occur. For our implementation, we opt for Python 3 as a general-purpose programming language widely used for data analysis \citep{vanRossum1997}. We aim to keep package dependencies to a minimum, making use of NumPy, SciPy, and pyts \citep[see][respectively]{Harris2020, Virtanen2020, Faouzi2020}, as well as the implementation of Equation~\ref{eq:vincenty} in GeoPy\footnote{\url{https://pypi.org/project/geopy}} for a fast computation of orthodromic distances.

Algorithm~\ref{alg:cluster} is the encompassing primary function, from which the others are called. After the spatial and temporal distance matrices are retrieved at the start, the algorithm iteratives through all data points, identifying suitable neighbors as described in Section~\ref{subsec:stdbscan}. If the latter does not occur, the respective points are designated as outliers. As the algorithm denotes a lack of cluster assignment as $-2$ and outlier status as $-1$, the last part of Algorithm~\ref{alg:cluster} shifts cluster numbers up by one to denote outliers as zero, with $\alpha_i \in \mathbb{N}_+$ and starting at one for cluster-assigned data points.

Algorithm~\ref{alg:spatial} calculates the spatial distance matrix as covered in Section~\ref{subsec:stdbscan}, with the density-based logistic rescaling introduced in Section~\ref{subsec:distance_rescaling}. The latter makes use of a kernel density estimate provided in Equation~\ref{eq:kde}, using Equation~\ref{eq:scott}, Equation~\ref{eq:silverman}, or a provided value for the bandwidth of the Gaussian kernel. Following this, Algorithm~\ref{alg:temporal} calculates the temporal distance matrix through dynamic time warping from Section~\ref{subsec:dtw}, using the DTW measure in Equation~\ref{eq:dtw}.

Finally, Algorithm~\ref{alg:nb} operates as a function to determine cluster-coherent neighbors by checking whether both spatial and temporal distances for data points and their corresponding time series fall below the provided distance threshold, following the traditional ST-DBSCAN algorithm's dual distance requirement.

The software implementation of SCADDA is available for download via the Python Package Index\footnote{(Final version will be added to the Python Package Index for the manuscript proof.)}, allowing for an easy installation through package managers. The software package can be installed with the command \texttt{`pip install scadda'}. Alternatively, the source code can also be downloaded from the project's GitHub repository\footnote{\label{ft:github_scadda}\url{https://github.com/moews/scadda}}, and the file \texttt{scadda.py} placed into the respective working directory.

Opting for a terminal installation is strongly recommended, as package requirements will be automatically checked and, if necessary, guaranteed by installing missing dependencies detected during the process. The GitHub repository also contains a basic code example, algorithm outputs, and the the corresponding dataset for the experiment using broadband antibiotics prescriptions in Scotland covered in Section~\ref{subsec:nhs_replication}.

\begin{minipage}[t]{0.49\linewidth}
\vspace{0pt}
\begin{algorithm}[H]
    \label{alg:cluster}
    \caption{Clustering function.}
	$\underline{\mathrm{CLUSTER}(\mathbf{s}, \mathbf{t}, \epsilon_s, \epsilon_t, \lambda, k, \xi):}$\\
	$\delta \leftarrow \mathrm{SPATIAL}(\mathbf{s}, k)$\\
	$\gamma \leftarrow \mathrm{TEMPORAL}(\mathbf{t}, \xi)$\\
	$c \leftarrow 0$\\
	$\tau \leftarrow 0$\\
	$\alpha \leftarrow -2_{|\delta|}$\\
	\For{$i \gets 0 \ \mathrm{to} \ |\delta|$}{
	    $\eta \leftarrow \mathrm{NB}(i, \delta, \gamma, \epsilon_s, \epsilon_t)$\\
	    \eIf{$|\eta| < \lambda$}{
	        $\alpha_i \leftarrow -1$
	    }{
	        $c \leftarrow c + 1$\\
	        $\alpha_i \leftarrow c$\\
	        \For{$j \gets \{ \eta \}$}{
	            $\alpha_j \leftarrow c$\\
	            $\tau \leftarrow \tau + 1$
	        }
	        \While{$|\tau| > 0$}{
	            $j \leftarrow \tau_{|\tau| - 1}$\\
	            $\tau \leftarrow \tau - 1$\\
	            $\eta_\mathrm{new} \leftarrow \mathrm{NB}(j, \delta, \gamma, \epsilon_s, \epsilon_t)$\\
	            \If{$|\eta_\mathrm{new}| > \lambda - 1$}{
	                \For{$l \gets \{ \eta_\mathrm{new} \} $}{
	                    $c_\mathrm{new} \leftarrow \alpha_{\eta_\mathrm{new}}$\\
	                    \If{$c_\mathrm{new} = -2$}{
	                        $\alpha_{c_\mathrm{new}} \leftarrow c$\\
	                        $\tau \leftarrow \tau + 1$
	                    }
	                }
	            }
	        }
	    }
    }
    $\alpha \leftarrow \alpha +_\mathrm{ele} 1$\\
    \For{$i \gets 0 \ \mathrm{to} \ |\alpha|$}{
        \If{$\alpha_i > 0$}{
            $\alpha_i \leftarrow \alpha_i - 1$
        }
    }
    \Return $\alpha$\\
	\vspace{10pt}
\end{algorithm}
\end{minipage}
\begin{minipage}[t]{.5\linewidth}
\vspace{0pt}
\begin{algorithm}[H]
    \label{alg:spatial}
    \caption{Spatial distance matrix.}
	$\underline{\mathrm{SPATIAL}(\mathbf{s}, k):}$\\
	$\psi \leftarrow \mathrm{KDE}_\mathrm{Gaussian} (\mathbf{s})$\\
	$\mu_\psi \leftarrow \frac{1}{|\mathbf{s}|} \sum_{i = 0}^{|\mathbf{s}|} \frac{\mathbf{s_i}}{|\mathbf{s}|}$\\
	$\delta \leftarrow 0_{|\mathbf{s}|, |\mathbf{s}|}$\\
	\For{$i \gets 0 \ \mathrm{to} \ |\mathbf{s}|$}{
	    \For{$j \gets i + 1 \ \mathrm{to} \ |\mathbf{s}|$}{
	        $w_1 \leftarrow \psi(\mathbf{s}_{i, \cdot})$\\
	        $w_2 \leftarrow \psi(\mathbf{s}_{j, \cdot})$\\
	        $\mu_w \leftarrow \frac{w_1 + w_2}{2}$\\
	        $w_\mathrm{log} \leftarrow \frac{2}{1 + e^{-k(\mu_w - \mu_\psi)}}$\\
	        $\delta_{i, j}, \delta_{j, i} \leftarrow w_\mathrm{log} \cdot \mathrm{D}(\mathbf{s}_{i, \cdot}, \mathbf{s}_{j, \cdot})$
	    }
	}
	\Return $\delta$\\
	\vspace{10pt}
\end{algorithm}
\begin{algorithm}[H]
    \label{alg:temporal}
    \caption{Temporal distance matrix.}
	$\underline{\mathrm{TEMPORAL}(\mathbf{t}, \xi):}$\\
	$\gamma \leftarrow 0_{|\mathbf{t}|, |\mathbf{t}|}$\\
	\For{$i \gets 0 \ \mathrm{to} \ |\mathbf{t|}$}{
	    \For{$j \gets i + 1 \ \mathrm{to} \ |\mathbf{t}|$}{
	        $\gamma_{i, j}, \gamma_{j, i} \leftarrow \mathrm{DTW}(\mathbf{t_{i, \cdot}}, \mathbf{t_{j, \cdot}}, \xi)$
	    }
	}
	\Return $\gamma$\\
	\vspace{10pt}
\end{algorithm}
\begin{algorithm}[H]
    \label{alg:nb}
    \caption{Neighbor identification.}
	$\underline{\mathrm{NB}(f, \delta, \gamma, \epsilon_s, \epsilon_t):}$\\
	$\eta \leftarrow \varnothing$\\
	\For{$i \gets 0 \ \mathrm{to} \ |\delta|$}{
	    \If{$i \neq f$}{
	        $d_1 \leftarrow \delta_{f, i}$\\
	        $d_2 \leftarrow \gamma_{f, i}$\\
	        \If{$d_1 < \epsilon_s \ \& \ d_2 < \epsilon_t$}{
	            $\eta \leftarrow \eta \cup \{ i \}$
	        }
	    }
	}
	\Return $\eta$\\
	\vspace{10pt}
\end{algorithm}
\vspace{10pt}
\end{minipage}


\section{Experiments and results}
\label{sec:experiments}

\subsection{Idealized validation experiment}
\label{subsec:toy_example}

As an initial testbed for our approach and implementation, we compare the spatial clustering component of SCADDA to that of ST-DBSCAN as the underlying percursor. The omission of temporal data is motivated by the equivalency between both algorithms in that regard, assuming that the latter makes use of dynamic time warping as well. In practical terms, this means that the spatial component of our implementation is compared to the traditional DBSCAN algorithm.

For this purpose, we draw 100 samples $X_{i, j} \sim \mathcal{N} (\mu_i, \sigma_j^2)$ each from eight different normal distributions, with means $\mu_i \in \{ (4, 4), (4, 8), (8, 4), (8, 8) \}$ and standard deviations $\sigma_j \in \{ 1.0, 0.1 \}$ on the diagonal of the respective covariance matrices, with  $\mathrm{cov} (X_1, X_2) = 0$ for independent variables. This presents us with four broader Gaussians, each of which has a second Gaussian with the same mean and a smaller variance, thus creating a density spike in each distribution's center.

\begin{figure}[!htb]
\centering
\includegraphics[width=\textwidth]{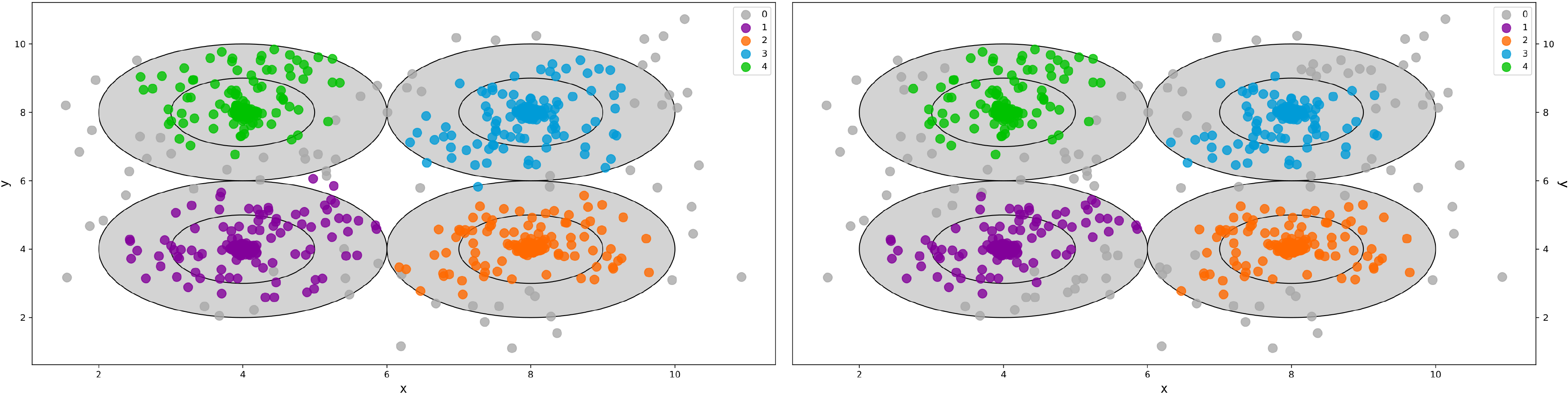}
\caption{Clustering of an idealized test case. The first and second panel from the left show results for the spatial components of SCADDA and ST-DBSCAN, respectively. The spatial data is comprised of eight Gaussians, with two each sharing a mean $\mu = \{ (4, 4), (4, 8), (8, 4), (8, 8) \}$ and differing in their standard deviation $\sigma = \{ 1.0, 0.1 \}$ to create high-density centers. Outliers are indicated with dark grey dots, while the first and second standard deviations of the broader Gaussians with $\sigma = 1.0$ are shown as black outlines shaded in light grey.}
\label{fig:toy}
\end{figure}

Figure~\ref{fig:toy} shows the first and second standard deviations of these enveloping broader Gaussians, in both panels, as black outlines shaded in light grey. Cluster assignments for clusters $c_i \in \{ 1, 2, 3, 4 \}$ are indicated as listed in the panels' legends, while outliers are depicted in dark grey. The left-hand panel shows the cluster assignments for SCADDA, whereas the right-hand panel shows those for DBSCAN.

We use the ST-DBSCAN functionality in our implementation to avoid any computational differences, making use of the same SCADDA implementation for both experiments. We achieve this purely spatial clustering process by creating step-invariant and identical dummy time series with $t_\mathrm{step} = 0$ for each time step, meaning that the temporal distance requirement is effectively waved through.

As expected, both algorithms find the four clusters built from two Gaussians each, with our approach outperforming the base ST-DBSCAN method by providing extended coverage for every one of them. While the latter leads to 13.88\% of data points being labeled as outliers, this is reduced to 9.12\% for SCADDA. The density-based rescaling of the spatial distance matrix proves useful when faced with high-density subgroups in clusters, which translates well to real-world datasets that feature, for example, urban versus rural areas.

\subsection{Replication of NHS Scotland results}
\label{subsec:nhs_replication}

Prescription volumes for antibiotics are known to feature seasonality \citep[see, for example,][]{Durkin2018}, as well as area-level drivers from socio-economic factors \citep{Molter2018}. Drug prescription patterns can, therefore, help to analyze both correlations with demographic variables and overprescription by general practitioners, and spatio-temporal modeling for related prescription correlation are established in the literature \citep{Blangiardo2016, Ashworth2021}.

Here, we recreate the experimental setup by \citet{Gieschen2022}, described in Section~\ref{sec:introduction}, using the data, analysis, and settings from their study leading to this work, to demonstrate the effect of highly variable population densities on clustering in healthcare analysis. We procure the same dataset from NHS Scotland publications and the Open Data Platform\footnote{\url{https://www.opendata.nhs.scot}} of the NHS Information Services Division.

The resulting dataset contains the aggregated prescription amount of medications based on amoxicillin, a common antibiotic, per general practitioner (GP) location and month, and covering the time frame from October 2015 to September 2017. GP locations are identified by unique practice codes and associated latitude and longitude coordinates. After dropping entries that feature missing coordinates, this results in 980 GP locations with corresponding prescription time series. This pre-processed dataset is also available as open-source data in the repository listed in Footnote~\ref{ft:github_scadda}.

\begin{figure}[!htb]
\centering
\includegraphics[width=\textwidth]{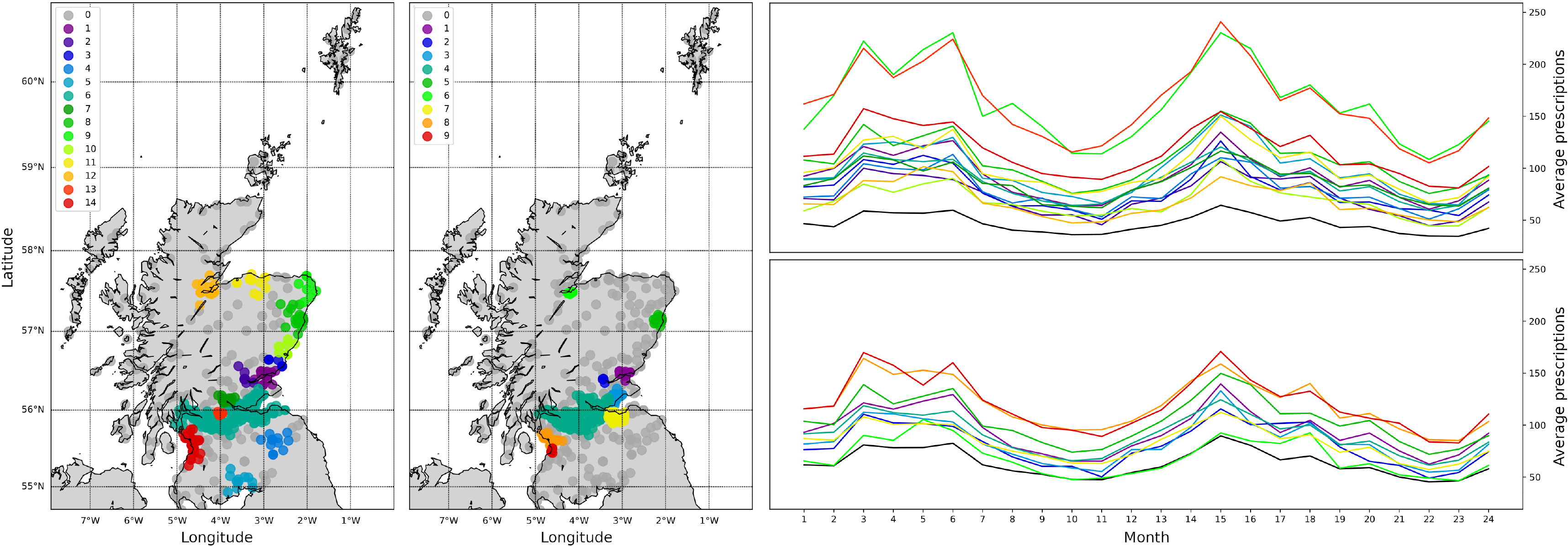}
\caption{Clustering of amoxicillin prescriptions in Scotland. The first and second panel from the left show clustering results using SCADDA and ST-DBSCAN, respectively, for spatial locations of general practitioners and their prescription behavior of the investigated antibiotic as monthly time series. The two horizontally stacked panels on the right-hand side show the averaged time series values per month and cluster, with results for SCADDA shown in the upper panel and results for ST-DBSCAN shown in the lower panel. Outliers are indicated with dark grey dots.}
\label{fig:scotland}
\end{figure}

The two vertically stacked panels on the right-hand side of Figure~\ref{fig:scotland} show these time series, averaged per identified cluster and month, with the upper and lower panel showing the results for SCADDA and ST-DBSCAN, respectively. For this experiment, we also test the application of z-score normalization of time series to ensure proper functionality. Notably, the time series feature considerable seasonality, with prescription peaking around December to January. This confirms other studies on seasonal effects and concentrations of antibiotics use in a visually accessible manner \citep{Sun2012, Minalu2013}.

The two left-hand panels show the locations of cluster-assigned GPs, for SCADDA and ST-DBSCAN in the first and second panel, respectively. As in \citet{Gieschen2022}, we note that our approach succeeds in combining closely associated clusters due to a stretching of distance weights in high-density areas. This is visible in the incorporation of clusters 3, 7 and 8 in the map on the right into clusters 6 and 14 in that on the left. This translates to the combination of the clusters for Edinburgh as well as Dunfermline and Kirkcaldy into the wider Central Lowlands cluster including Glasgow for the first case, while the second case corresponds to the combination of the cluster around Ayr with the one covering Irvine and Kilmarnock.

While the right-hand map is only able to identify narrow clusters around Inverness and Aberdeen, the left-hand map broadens these, and also manages to retrieve cluster assignments for Lossiemoth, Peterhead and Fraserburgh, Arbroath, Stonehaven and Montrose, the wider Dumfries and Galloway area, and the Scottish Borders. Clusters for both Stirling and the lower-density area in North Lanarkshire, in green and light red, respectively, on the other hand, are split off from the central urban cluster.

The corresponding time series for the two methods also demonstrate a starker differentiation, with the vertical scale held constant between the lower and upper right-hand panels. Notably, the two high-prescription series coincide with two of the socioeconomically most disadvantaged spots according to the Scottish Index of Multiple Deprivation\footnote{\url{https://simd.scot}} (SIMD) in North Lanarkshire and around Peterhead, with the latter also being the UK's largest fishing port, which could further exacerbate the need for antibiotic medications through occupational hazards and a high flux of people passing through the port. In both panels, black lines indicate outliers, showing that these cover areas with lower average prescription numbers.

\subsection{GFED4 carbon emission from fires}
\label{subsec:case_study}

Fires are a major driver of the release of greenhouse gases and aerosols, contributing to atmospheric pollution and climate change. The primary source of burned area are fires in Sub-Saharan Africa, which is estimated to contribute around 70\% of global burned area and 50\% of carbon emissions from fires, respectively \citep{Andela2014, Ramo2021}. Modelling fires and their spatial and temporal variability well is crucial for predicting biogeochemical cycling, pyrogenic emissions and vegetations patterns, where carbon emissions in particular are highly relevant for global circulation models \citep{Lehsten2010}.

The collection of data relies on Earth-observing satellites, for example through the Pathfinder mission jointly between the National Aeronautics and Space Administration (NASA) and the National Agency of National Oceanic and Atmospheric Administration (NOAA), as well as through the Climate Change Initiative of the European Space Agency (ESA) \citep{Riano2007, Chuvieco2016}.

To study the application of our spatio-temporal clustering approach to the spatial distribution of burned area and carbon emissions over time, we make use of the fourth-generation data release of the Global Fire Emissions Database\footnote{\url{https://www.globalfiredata.org}} (GFED4) \citep{Randerson2012, Giglio2013, vanderWerf2017}.

\begin{figure}[!htb]
\centering
\includegraphics[width=\textwidth]{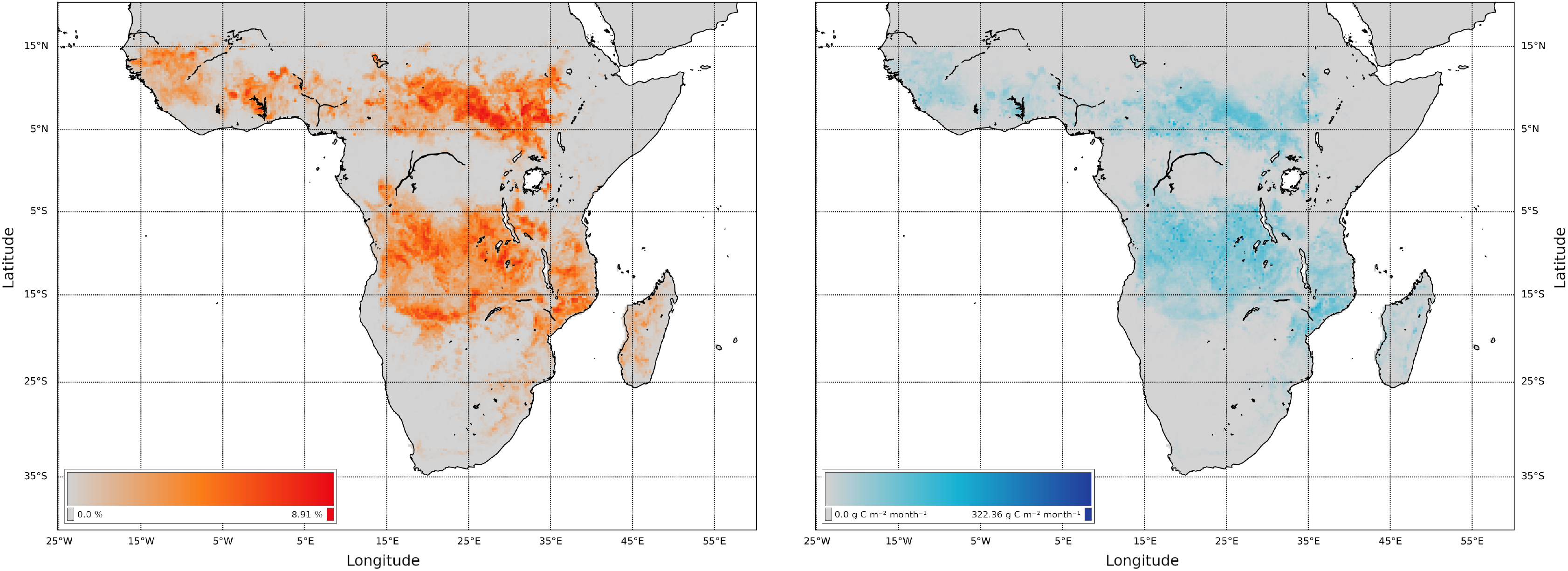}
\caption{Density maps for burned area fractions and carbon emissions for Sub-Saharan South Africa. Data underlying the figure are extracted from the fourth-generation Global Fire Emissions Database (GFED4). The left-hand panel shows the average burned area as the fraction per resolution area from 2012 to 2016, from zero to a maximum of 8.91\%. The right-hand panel shows average fire carbon emissions during the same time frame, covering 60 separate months of observations, from zero to a maximum of 322.36 grams of carbon per square meter per month.}
\label{fig:africa_1}
\end{figure}

We extract GFED4 data on burned area and carbon emissions from the beginning of 2012 to the end of 2016, as later years are still in the beta release phase. We then average the burned area over this time period, and create separate monthly data slices for carbon emissions, leading to a time series of length $n = 60$ per resolution area. Figure~\ref{fig:africa_1} shows, in the left-hand panel, burned area as the fraction of resolution areas, while the right-hand panel shows fire carbon emissions in grams of carbon per square meter and month.

The low-density band following the equator correlates with the presence of influencing factors. The bare spot in Central Africa around the Congo River in particular coincides with tropical rain forests, which also feature increased humidity and precipitation. Low-density areas in terms of burned area in North Africa and South Africa, on the other hand, are low in humidity, but are also the home to vast stretches of deserts and semi-deserts. The high-density areas above and below the equator, as well as Madagascar, overlap with the distribution of dry forests on the continent, spanning from deciduous forests to savannas and shrublands \citep[see, for example,][for a comparison]{Miles2006}. 

We standardize the latitude-longitude matrix of burned area fractions $B$, shown in the left-hand panel of Figure~\ref{fig:africa_1}, to probabilities, here denoted as $\hat{B}$ for simplicity, so that $\sum_{i = 1}^{|B|} \sum_{j = 1}^{|B|} \hat{B}_{i, j} = 1$. We can then draw coordinate samples with these probabilities, essentially employing the resampling step of a sampling importance resampling approach, to transform a matrix of density values per grid tile into latitude-longitude samples representing the underlying density distribution. We use an empirical heuristic,
\begin{equation}
\epsilon_s = \left( \frac{\sigma (\mathbf{s}_{\cdot, 1}) + \sigma(\mathbf{s}_{\cdot, 2})}{2} \right) \left( \frac{\max(\mathbf{s}_{\cdot, 1}) - \min(\mathbf{s}_{\cdot, 1})}{2} + \frac{\max(\mathbf{s}_{\cdot, 2}) - \min(\mathbf{s}_{\cdot, 2})}{2} \right)^{-2},
\label{eq:thumb_2}
\end{equation}
which stems from experiments in Section~\ref{subsec:toy_example} and which we will discuss further in Section~\ref{sec:discussion}, and use the same arithmetic mean of temporal distance matrix values as for the NHS Scotland validation in Section~\ref{subsec:nhs_replication},
\begin{equation}
\epsilon_t = |\mathbf{t}|^{-2} \cdot \sum_{i = 1}^{|t|} \sum_{j = 1}^{|t|} \mathrm{DTW} (\mathbf{t}_{i, \cdot}, \mathbf{t}_{j, \cdot}, \xi),
\label{eq:thumb_3}
\end{equation}
for the temporal distance limit. In order to showcase the impact that time series data concerning carbon emissions have on the cluster identification, we perform the clustering in this experiments twice; once using only spatial information on burned area fractions, by setting uniform dummy time series values at zero, and once with the addition of monthly fire carbon emission data.

\begin{figure}[!htb]
\centering
\includegraphics[width=\textwidth]{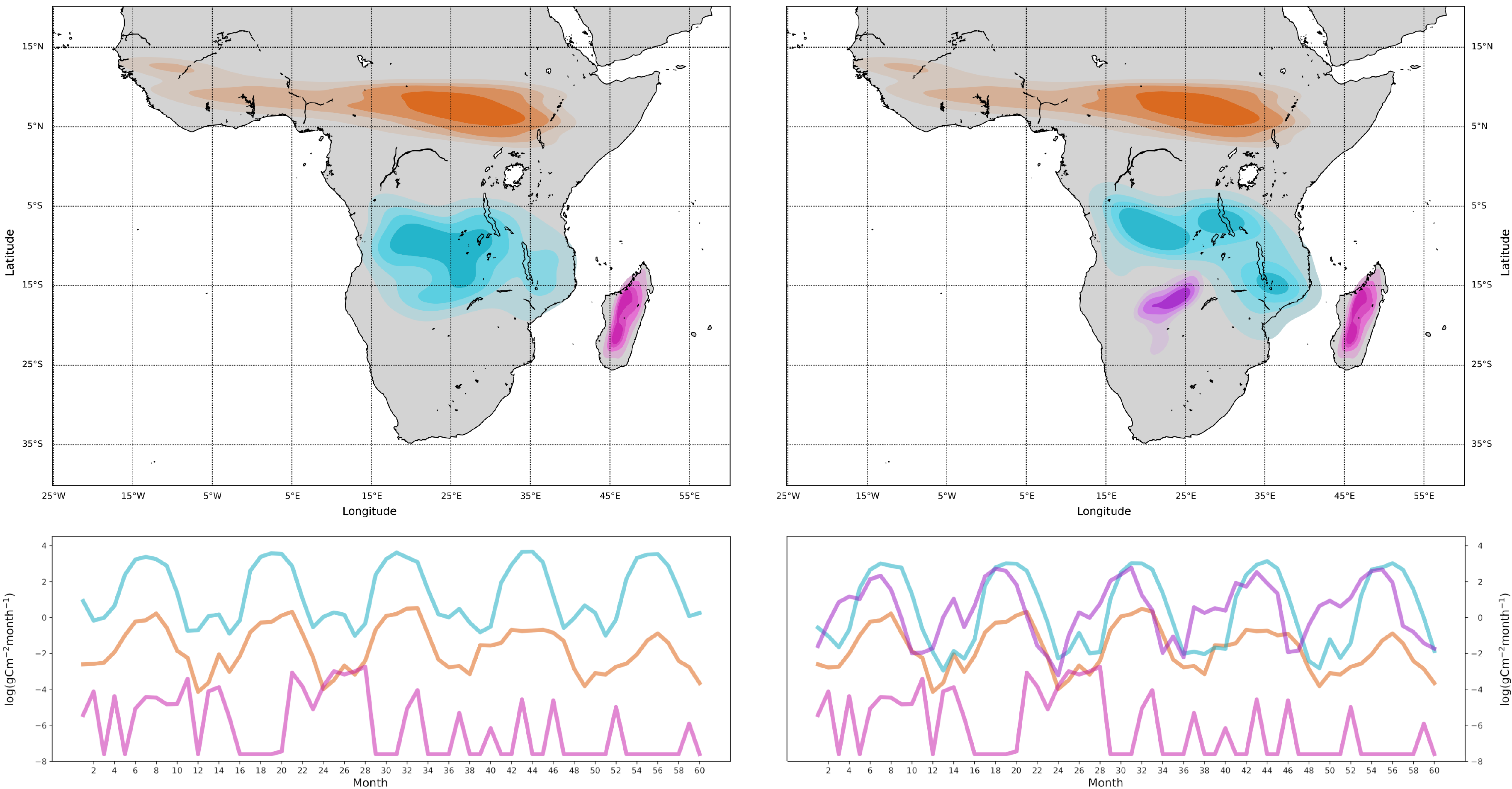}
\caption{Clustering of burned area densities and carbon emissions in Sub-Saharan Africa. Data underlying the figure are extracted from the fourth-generation Global Fire Emissions Database (GFED4). The two upper panels show color-coded kernel density estimates of identified clusters, with the left-hand and right-hand panels depicting the results for a purely spatial burned area analysis and a spatio-temporal clustering with additional monthly carbon emissions, respectively. The two lower panels show average monthly carbon emission values per cluster, in grams of carbon per square meter and month.}
\label{fig:africa_2}
\end{figure}

The results for the former and latter are shown in the upper-left and upper-right panels of Figure~\ref{fig:africa_2}, respectivly, with the lower panels showing the corresponding average carbon emissions per month and cluster.

As the spatial data points are density-sampled rather than empirical locations such as in the NHS Scotland experiment in Section~\ref{subsec:nhs_replication}, we apply a KDE to the points for each cluster assignment to recapture the density distribution in a visually accessible manner, and plot each cluster in a different color. The two primary identified clusters for the spatial-only experiment using burned area data correspond to the two bands around 10$^{\circ}$N, approximately from Guinea to Ethiopia, and around 10$^{\circ}$S, approximately from Angola to Mozambique. The associated time series show a clear seasonality, with peaks and valles being identificable for both the northern-hemisphere cluster and the southern-hemisphere cluster.

The time series for Madagascar in pink demonstrates considerably less seasonality, although this is rooted in both the analysis and environmental factors. The sample size, due to a lower overall density relative to the continent as a whole, means that our sampling process does not extract sufficient sample sizes to retrieve reasonably smooth measurements. The second factor is due to the broad range of ecological regions across the comparatively small area of the island, which leads to shifted seasonal effects per ecological region as described in detail by \citet{FrappierBrinton2022}. This means that different types of environments overlap in their fire-based carbon emissions, in addition to the widespread practice of slash-and-burn agriculture \citep{Styger2007}.

The inclusion of temporal data on carbon emissions in the right-hand panels showcases the importance of spatio-temporal clustering when trying to analyze seasonality as well as distributions of fires and associated emissions. The prior southern-hemisphere cluster distinguishes more clearly between peaks in central Mozambique, the southern border of Tanzania and the Democratic Republic of the Congo, and the border region of the latter with Angola. The elongated density peak in the border region between Angola, Namibia, Zambia, Botswana, and Zimbabwe that is visible in the left-hand panel of Figure~\ref{fig:africa_1} is separated from the cluster. The reason is reflected in the corresponding time series, with a backward seasonality shift. The remaining two clusters stay identical, demonstrating the stability of the clustering process.


\section{Discussion}
\label{sec:discussion}

As the applicability of different clustering methods depends on the problem at hand, our approach and implementation targets a scenario which arises in geospatial analysis and consists of two qualifiers. The first is the need for a spatio-temporal approach due to both spatial coordinate values and associated time series data, although providing dummy time series of zero-filled vectors can reduce our algorithm to a purely spatial application, as demonstrated in Sections~\ref{subsec:toy_example} and~\ref{subsec:case_study}. Our implementation is geared toward geospatial applications, which means that it deals with two-dimensional spatial coordinates, but accepts any time series length. The second is the need to amend the clustering process due to highly variable spatial densities as an extraneous variable, leading to a desire to soften the impact of these densities, for example when urban versus rural population densities would otherwise constrain the cluster assignment too heavily.

SCADDA inherits several advantages of the base DBSCAN algorithm, namely the lack of a requirement to set a number of clusters in advance, the ability to identify arbitrarily-shaped clusters, and the robustness to outliers. The density-aware distance rescaling also does away with a major disadvantage of the base algorithm, which is the inability to deal with large differences in densities with a global spatial distance limit \citep{Kriegel2011}. For this purpose, we extend the approach taken by \citet{Gieschen2022} for rescaling with a bounded logistic function, as well as their application of dynamic time warping to the ST-DBSCAN algorithm as the spatio-temporal extension of the base method. As a result, our implementation is well-suited for geospatial clustering challenges featuring temporal data per coordinate, with a need for accurate distance metrics, allowing for irregularly-shaped clusters, and circumventing the problems arising from large densitiy differences in prior algorithms.

The use of KDE, which is employed to determine the local density of two data points between which a distance is calculated, presents a limitation. On the one hand, the choice of kernel is a constraint put on the respective dataset, although Gaussian kernels are a common choice for these kinds of analyses. In addition, the bandwidth of the kernel, in our case the standard deviation of said normal distribution, is set globally, while \citet{Shi2010} argue in favour of adaptive bandwidths for select geospatial analyses. Here, research efforts for locally-determined bandwidths are part of a continuing effort to optimize density estimates \citep[see, for example,][]{Farmen1999, Sadiq2022}, and implementing a secondary scaling process by incorporating these works is an interesting methodological avenue for future work.

Concerning bandwidth computations more broadly, \citet{Filippone2011} provide an approximate Bayesian bandwidth estimate using the expectation-propagation algorithm, whereas \citet{Obrien2016} extend an earlier method by \citet{Bernacchia2011} from the univariate to the multivariate case. That being said, this could, for the type of extremely heterogeneous density distributions that our approach is designed for, lead to spike-like sets of narrow kernels in, for example, city centers. If carefully applied, however, this potentially offers an additional way to fine-tune the method to specific datasets. At the same time, the computational cost associated with elaborate bandwidth approximations must be balanced against the sufficiency of established heuristics in cases where these estimates are only a precursory step to provide a general density landscape for distance rescaling. Another avenue is the convex combination of spatial and temporal distance matrices as shown in \citet{Deb2023}.

The spatial and temporal distance limits present another direction for further research, as Equations~\ref{eq:thumb_2} and~\ref{eq:thumb_3} are merely empirical heuristics that work well for investigated cases. The abovementioned rescaling alleviates the primary issue of DBSCAN-type algorithms in terms of a global spatial limit, but our method suffers from the same need to manually set limits as other algorithms in this family. Aside from $k$-distance plots, which can serve as a visual aid to determine these values by plotting the number of reachable neighbors versus different limits, a hierarchical alternative is given by the OPTICS algorithm, which linearly orders data points according to their spatial distance \citep{Ankerst1999}. While beyond the scope of this work, a methodologically oriented study on these kinds of alternatives could prove fruitful to contribute to distance limits in spatial clustering algorithms more broadly.

From a computational standpoint, we intend to follow up on our work by integrating parallel computing capabilities into our implementation. This applies to the creation of distance matrices in particular, as they are inherently embarrassingly parallel due to the independence of each entry's calculation. The projected use cases for this extension are research questions dealing with very fine-grained structures and strong accuracy requirements, which necessitates larger samples. Here, the `schwimmbad' package by \citet{PriceWhelan2017} offers an accessible way to include both single-machine multiprocessing abilities and larger-scale pools for supercomputing infrastructures.

Lastly, we want to discuss extensions of our presented application case and further potential domain applications. Staying within the realm of fire carbon emissions, we intend to collaborate with domain experts to go beyond the novel insights into cluster separation based on carbon emission time series in different areas of Sub-Saharan Africa. Follow-up researc in this context can be split into two categories; one that deals with more detailed insights into emissions on the investigated continent and relations to climate change patterns as a more direct continuation, and one that transfers these developments to current challenges in wildfire analysis in North America. An additional pathway is the extension to additional application cases, straying beyond health and environmental statistics, for example to the social sciences through a transfer to our prior work on criminological incident reports \citep{Moews2021}.


\section{Conclusion}
\label{sec:conclusion}

In this paper, we introduce a method and implementation for spatio-temporal clustering when addressing geospatial data with highly variable densities as an extraneous factor. The contributions of this paper, which is centered on a novel computational tool, cover methodology, software, and domain applications.

We build on the ST-DBSCAN algorithm and employ prior research that introduces the concept of continuous density-aware distance rescaling as well as the use of dynamic time warping to the former algorithm. We extend this approach by applying the Sakoe-Chiba band to the efficiency of dynamic time warping, adding an option for z-score normalization of time series data, and introducing a parameter for a maximum percentage of outliers into the algorithm. We also update the spatial distance metric to a choice between computationally less expensive and accurate alternatives that avoid rounding errors in special cases, and include a more in-depth coverage of the underlying statistical methods.

Our work presents the, to our knowledge, first open-source implementation of ST-DBSCAN with dynamic time warping at the time of writing, as well as the first publicly available software tool incorporating density-aware distance rescaling. With a high degree of customizability for domain applications in mind, the resulting software allows for a multitude of parameters to be set, and automatically defaults to best practices and established rules. The implementation also features a regular ST-DBSCAN option without distance rescaling, and extensive parameter checks to ensure no runtime errors.

For suitable domain applications of interest, and after confirming the algorithm's functionality on a simple test case of nested multimodal distributions, we first reconstruct prior work and find sensible clusterings for antibiotics prescriptions in Scotland that follow socioeconomic demographics. We then test our approach on burned area fractions and carbon emissions from fire radiative power observations drawn from high-resolution spectroradiometers onboard NASA's two Terra and Aqua satellites, which to our knowledge is the first application of DBSCAN-type algorithms to spatio-temporal carbon emissions in Africa.

We retrieve the two primary clusters of fires north and south of the equatorial rain forest band, which correlate to less humid dry forest regions, as well as Madagascar, and confirm strong seasonality. When using temporal data on carbon emissions, we also separate a cluster in central Southern Africa, which features shifted seasonality peaks, demonstrating the impact of combined spatio-temporal approaches. Our findings contribute to the literature on fires and carbon emissions for ecological planning and climate science, as well as more broadly to the available methodology and accessibility of spatio-temporal clustering approaches.

\section*{Acknowledgments}

Our thanks go to Johann Faouzi and Hicham Janati for accessible work on open-source dynamic time warping as well as the multitude of contributors of the SciPy, NumPy, and GeoPy frameworks. We also wish to acknowledge the broad literature on geodesy and geospatial statistics that we were able to draw from in the process of completing this work. The interdisciplinarity involved does not go quite as far as to unite the (academic) world, but at least lets us measure it together.

\section*{Declarations of interest and funding}

Declarations of interest: None. This research did not receive any specific grant from funding agencies in the public, commercial, or not-for-profit sectors. 

\section*{References}

\bibliographystyle{apalike}
\bibliography{ref.bib}

\end{document}